\documentclass[twocolumn,twoside,slac_two]{revtex4}
\usepackage{graphicx}
\usepackage{fancyhdr}
\usepackage{hyperref}
\hypersetup{
    colorlinks=true,
    urlcolor=magenta,
}
\newcommand{\rom}[1]{\textsc\expandafter{\romannumeral #1\relax}}
\usepackage{mciteplus}

\begin{document}

\title{Progress Report on the Measurement of the Carbon Flux in Cosmic Rays from 1.9 GV to 2.6 TV with the Alpha Magnetic Spectrometer on the International Space Station}

%

\author{Yang Li, on behalf of the AMS Collaboration}
\affiliation{DPNC, Universit\'{e} de Gen\`{e}ve, CH–1211 Gen\`{e}ve 4, Switzerland}
\affiliation{Sun Yat-Sen University (SYSU), Guangzhou, 510275, China}

\begin{abstract}
Knowledge of the precise rigidity dependence of the carbon flux is important in understanding the origin, acceleration and propagation of cosmic rays. The current status of the measurement of the carbon flux in cosmic rays with rigidity from 1.9 GV to 2.6 TV, based on 8.3 million events collected by the Alpha Magnetic Spectrometer during the first 5 years of operation, is presented.
\end{abstract}

\maketitle

\thispagestyle{fancy}


\section{Introduction}
Over the last 30 years there have been many measurements of the carbon flux in cosmic rays (CRs) by balloon and satellite experiments~\cite{bib:Panov2009, *bib:Juliusson1974, *bib:Orth1978, *bib:Lezniak1978, *bib:Derrickson1992, *bib:Simon1980, *bib:Buckley1994, *bib:Ahn2009, *bib:Mueller1991, *bib:Engelmann1990, *bib:Adriani2014, *bib:Obermeier2011}. The exact behavior of the carbon flux with rigidity is important in understanding the origin, acceleration and propagation of CRs. Here we report the current status of the carbon flux analysis in the rigidity range from 1.9 GV to 2.6 TV based on data collected by the Alpha Magnetic Spectrometer (AMS) during the first 5 years (May 19, 2011 to May 26, 2016) of operation onboard the International Space Station (ISS).

\section{Detector}
The layout and description of the AMS detector are presented in Ref.~\cite{bib:ams-02}. The key elements used in this measurement are the permanent magnet~\cite{bib:ams-mag}, the silicon tracker, four planes of time of flight (TOF) scintillation counters, and an array of 16 anticoincidence counters. The AMS also contains a transition radiation detector (TRD), a ring imaging \v{C}erenkov detector (RICH), and an electromagnetic calorimeter (ECAL).

The tracker~\cite{bib:ams-trk} has nine layers, the first (L1) at the top of the detector, the second (L2) above the magnet, six (L3 to L8) within the bore of the magnet, and the last (L9) above the ECAL. L2 to L8 constitute the inner tracker.

Together, the tracker and the magnet measure the rigidity $R$ of charged CRs. For carbon, the spatial resolution in each tracker layer is 10 $\mu$m on average in the bending direction and the resulting maximum detectable rigidity (MDR) is 2.6 TV over the 3 m lever arm from L1 to L9. Each layer of the tracker also provides an independent measurement of the charge $Z$ with a resolution of $\Delta Z/Z=5\%$ for carbon. Overall the inner tracker has a resolution of $\Delta Z/Z=2\%$ for carbon.

Two planes of TOF counters~\cite[and references therein]{bib:ams-tof} are located above L2 (upper TOF) and two planes are located below the magnet (lower TOF). The overall velocity ($\beta=v/c$) resolution has been measured to be $\Delta\beta/\beta^{2}=0.01$ for carbon nuclei. This discriminates between upward- and downward-going particles. The pulse heights of the two upper layers are combined to provide an independent measurement of the charge with an accuracy of $\Delta Z/Z=3\%$ for carbon. The pulse heights from the two lower planes are combined to provide another independent charge measurement with the same accuracy.

Carbon nuclei traversing AMS were triggered as described in detail in Ref.~\cite{bib:ams-he}. The trigger efficiency for carbon nuclei was measured to be $>98\%$ over the entire rigidity range.

Monte Carlo (MC) simulated events were produced using a dedicated program developed by the collaboration based on the {\sc Geant-4.10.1} package~\cite{bib:Geant4-1, *bib:Geant4-2}. The program simulates electromagnetic and hadronic interactions of particles in the AMS material and generates detector responses. The Glauber-Gribov model was used for the description of the inelastic cross-sections. The {\sc incl++} package~\cite{bib:INCL-1, *bib:INCL-2} was used to model nucleus-nucleus inelastic interactions below 5 GeV/nucleon and the {\sc dpmjet-\rom{2}.5} package~\cite{bib:DPMJET} was used at higher energies. The nucleus-nucleus hadronic elastic and quasi-elastic scatterings were treated using the model developed for helium nuclei~\cite{bib:ams-he} and validated by detailed comparison between data and simulation.

\section{Event Selection}
In the first 5 years the AMS collected $8.5\times10^{10}$ CR events. The collection time used in this analysis includes only those seconds during which the detector was in normal operating conditions and, in addition, AMS was pointing within $40^{\circ}$ of the local zenith and the ISS was outside of the South Atlantic Anomaly. Because of the influence of the geomagnetic field, this collection time for galactic CRs increases with rigidity becoming constant at $1.23\times10^{8}$ seconds above 30 GV.

Events are required to be downward going and to have a reconstructed track in the inner tracker and passing through L1. In the highest rigidity region, $R\ge0.88$ TV, the track is also required to pass through L9. Track fitting quality criteria such as a $\chi^{2}/\rm{d.f.}<10$ in the bending coordinate are applied, similar to Ref.~\cite{bib:ams-p, bib:ams-he}.

The measured rigidity is required to be greater than a factor of 1.2 times the maximum geomagnetic cutoff within the AMS field of view. The cutoff was calculated by backtracing~\cite{bib:backtr} particles from the top of the AMS out to 50 Earth's radii using the most recent IGRF~\cite{bib:IGRF-1, *bib:IGRF-2} geomagnetic model.

\begin{figure}[!h]
\centering
    \includegraphics[width=0.5\textwidth]{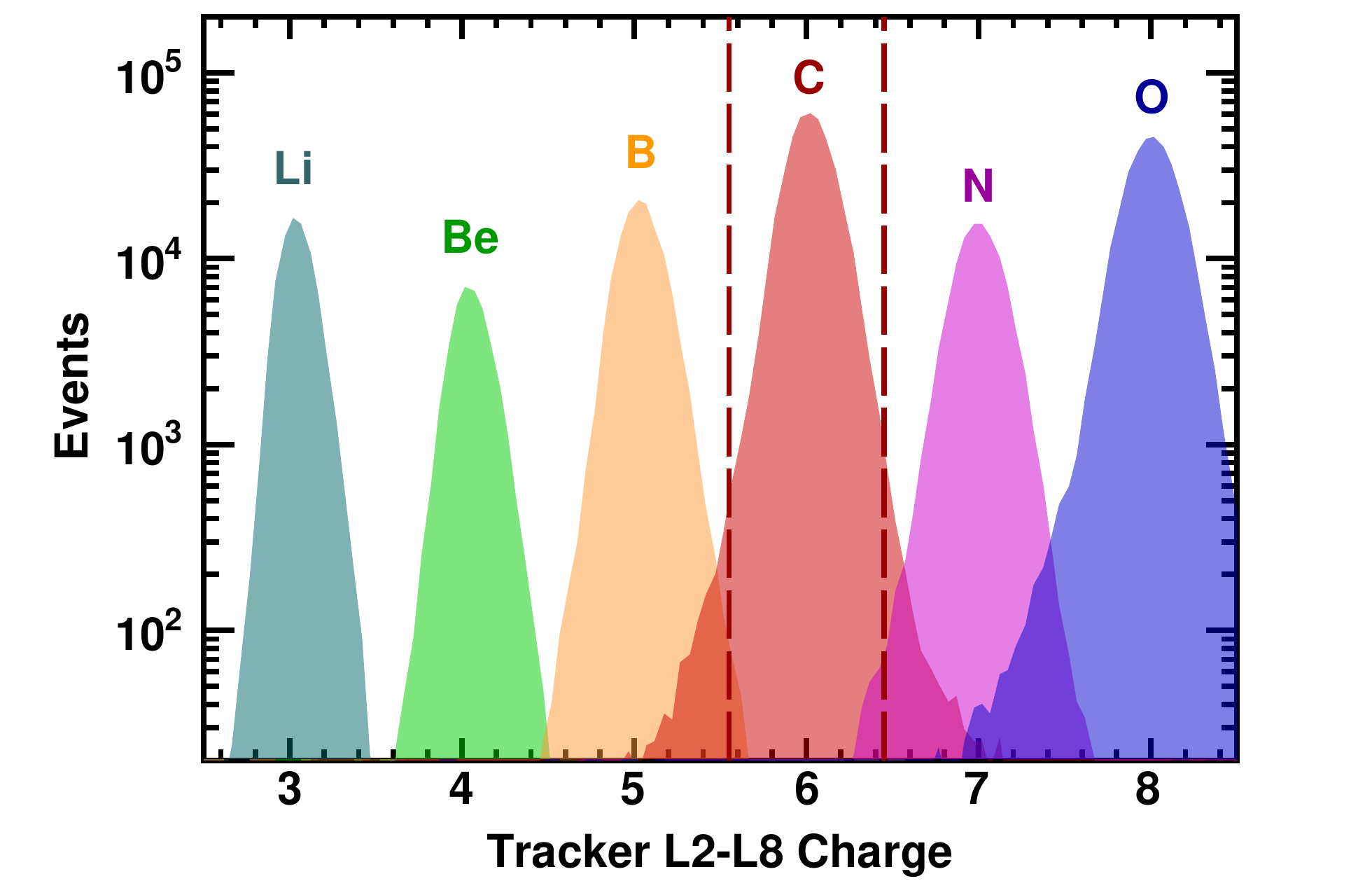}
\caption{Charge distribution of inner (L2--L8) tracker for samples from $Z=3$ to $Z=8$ in the rigidity range from 4 GV to 10 GV, selected by charge on L1, upper TOF, and lower TOF. The vertical dashed lines correspond to the charge selection in the inner tracker for carbon (red).}
\label{fig:fig1}
\end{figure}

Charge measurements on L1, inner tracker, upper TOF, lower TOF, and, for $R\ge0.88$ TV, L9 are required to be compatible with $Z=6$, as shown in Fig.~\ref{fig:fig1} for the inner tracker. This selection yields a purity of 99\% for carbon.

The residual background to carbon events resulting from interactions of heavy nuclei such as nitrogen and oxygen in the material between L1 and L2 (TRD and upper TOF) is evaluated by fitting the charge distribution of L1 with charge distribution templates of C, N, and O as shown in Fig.~2(b) in the Supplemental Material of Ref.~\cite{bib:ams-bc}. The charge distributions templates are obtained from a selection of non-interacting samples on L2 by the use of the charge measurements with L1 and L3--L8. This residual background is $<0.5\%$ for carbon sample.

The background from nitrogen and oxygen interactions on materials above L1 (thin support structures made by carbon fiber and aluminum honeycomb) has been estimated from simulation, using MC samples generated according to AMS flux measurements~\cite{bib:ams-nuclei}. The simulation of nuclear interactions has been validated using data. The background from interactions above L1 in the carbon sample is $<0.5\%$ over the entire rigidity range. After background subtraction the sample contains $8.3\times10^{6}$ carbon nuclei.

\section{Data Analysis}
The isotropic carbon flux $\Phi_i$ for the $i$th rigidity bin $(R_i, R_i+\Delta R_i)$ is
\begin{equation}\label{eq:flux}
\Phi_i={N_i\over A_i\, \epsilon_i\, T_i\, \Delta R_i},
\end{equation}
where $N_i$ is the number of events corrected for bin-to-bin migrations, $A_i$ is the effective acceptance, $\epsilon_i$ is the trigger efficiency, and $T_i$ is the collection time. In this analysis the carbon flux was measured in 67 bins from 1.9 GV to 2.6 TV with bin widths chosen according to the rigidity resolution.

The bin-to-bin migration of events was corrected using the unfolding procedure described in Ref.~\cite{bib:ams-p}.

Extensive studies were made of the systematic errors. These errors include the uncertainties in the two background estimations (below and above L1) discussed above, in the trigger efficiency, in the acceptance calculation, in the rigidity resolution function, and in the absolute rigidity scale.

The systematic error on the carbon flux associated with the trigger efficiency is $<0.5\%$ over the entire rigidity range.

The effective acceptances $A_i$ were calculated using MC simulation, and corrected for small differences between the data and MC simulation related to a) event reconstruction and selection, namely in the efficiencies of velocity determination, track finding, charge determination, and tracker quality cuts and b) the inelastic interactions in the AMS materials.

The systematic error on the carbon flux associated with the reconstruction and selection is $<1\%$ over the entire rigidity range.

\begin{figure}[!h]
\centering
    \includegraphics[width=0.5\textwidth]{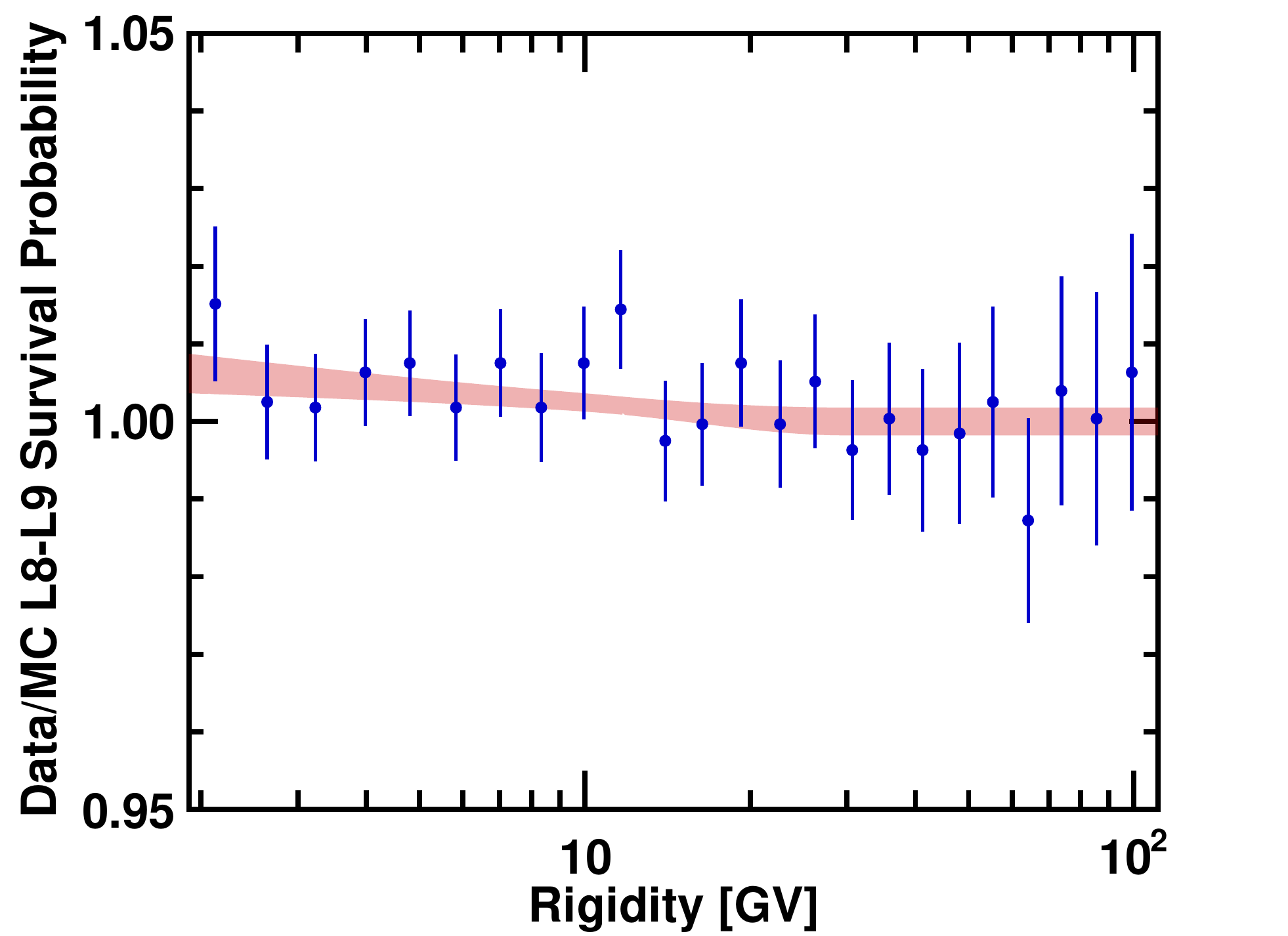}
\caption{The Data/MC ratio of carbon survival probabilities between tracker L8 and L9. The shaded areas show the systematic error range (68\% CL) obtained from the spline fit to data points.}
\label{fig:fig2}
\end{figure}

The material traversed by nuclei between L1 and L9 is composed primarily of carbon and aluminum, as described in detail in Ref.~\cite{bib:ams-he}. The corresponding inelastic cross-sections for C+C and C+Al have only been measured below 10 GV~\cite{bib:Cxsec-1, *bib:Cxsec-2, *bib:Cxsec-3, *bib:Cxsec-4, *bib:Cxsec-5, *bib:Cxsec-6, *bib:Cxsec-7, *bib:Cxsec-8, *bib:Cxsec-9}. To verify the MC predictions the carbon event samples that traverse materials between L8 and L9 (lower TOF and RICH) without interacting are measured in data and compared with MC simulations with Glauber-Gribov inelastic cross-sections varied within $\pm10\%$. The resulting cross-sections with the best agreement to data above 30 GV were chosen. Figure~\ref{fig:fig2} shows the ratio between the data and simulation of the carbon survival probabilities from L8 to L9. The survival probabilities between L1 and L2 have been calculated using data periods in which AMS was horizontal, i.e., $\sim90^{\circ}$ with respect to the zenith~\cite{bib:ams-he}. This independently verifies the inelastic cross-sections. The systematic error on the flux from the acceptances due to uncertainties of inelastic cross-sections is evaluated to be 2\% below 10 GV increasing to $3\%$ at 2.6 TV.

The rigidity resolution function $\Delta(1/R)$ for carbon is similar to that of the helium discussed in Ref.~\cite{bib:ams-he}. It has a pronounced Gaussian core characterized by width $\sigma$ and non-Gaussian tails more than $2.5\sigma$ away from the center. The rigidity resolution function was verified with four procedures. First, the differences of the coordinates measured in L3 or L5 to those obtained from the track fit using the measurements from L1, L2, L4, L6, L7 and L8 were compared between data and simulation. This procedure directly measures the tracker bending coordinate accuracy of 10 $\mu$m as shown in Fig.~5(b) in the Supplemental Material of Ref.~\cite{bib:ams-bc}. Similar results were obtained for the rest of the inner tracker layers. Second, the distribution of the scattering angle, defined as the difference between the inner tracker track and the L1 to L2 trajectory, was compared between data and simulation as shown in Fig.~6 in the Supplemental Material of Ref.~\cite{bib:ams-bc}, and found to be in good agreement. This comparison verifies the multiple, nucleus-nucleus elastic, and quasi-elastic scatterings. Third, in order to validate the alignment of the external layers L1 and L9, the difference between the rigidities measured using the information from L1 to L8 and from L2 to L9 was compared between data and simulation and found to be in good agreement similar to Ref.~\cite{bib:ams-p}. Fourth, the RICH velocity resolution is $\Delta\beta/\beta=5\times10^{-4}$~\cite{bib:ams-rich} for carbon. The rigidity resolution function up to 20 GV, including non-Gaussian tails, was obtained with data using the RICH velocity measurements only and found to be in good agreement with the rigidity resolution function from the simulation similar to Ref.~\cite{bib:ams-he}.

The first three verifications provided the MDR of 2.6 TV with 5\% uncertainty. The second, third and fourth verified the nucleus-nucleus elastic and quasi-elastic scattering in the AMS materials and provided the uncertainty of the amplitude of the non-Gaussian tails in the rigidity resolution function to be 10\%.

The systematic error on the carbon flux due to uncertainties in the rigidity resolution function was obtained by repeating the unfolding procedure varying the width of the Gaussian core of the resolution function by 5\% and the amplitude of the non-Gaussian tails by 10\% and found to be less than 1\% below 100 GV and 3\% at 2.6 TV.

There are two contributions to the systematic uncertainty on the rigidity scale, one from residual tracker misalignment and the other from magnetic field temperature corrections, as discussed in Ref.~\cite{bib:ams-p}. The corresponding error on the carbon flux is below $1\%$ up to 100 GV and reaches 6\% at 2.6 TV.

\begin{figure}[!h]
\centering
    \includegraphics[width=0.5\textwidth]{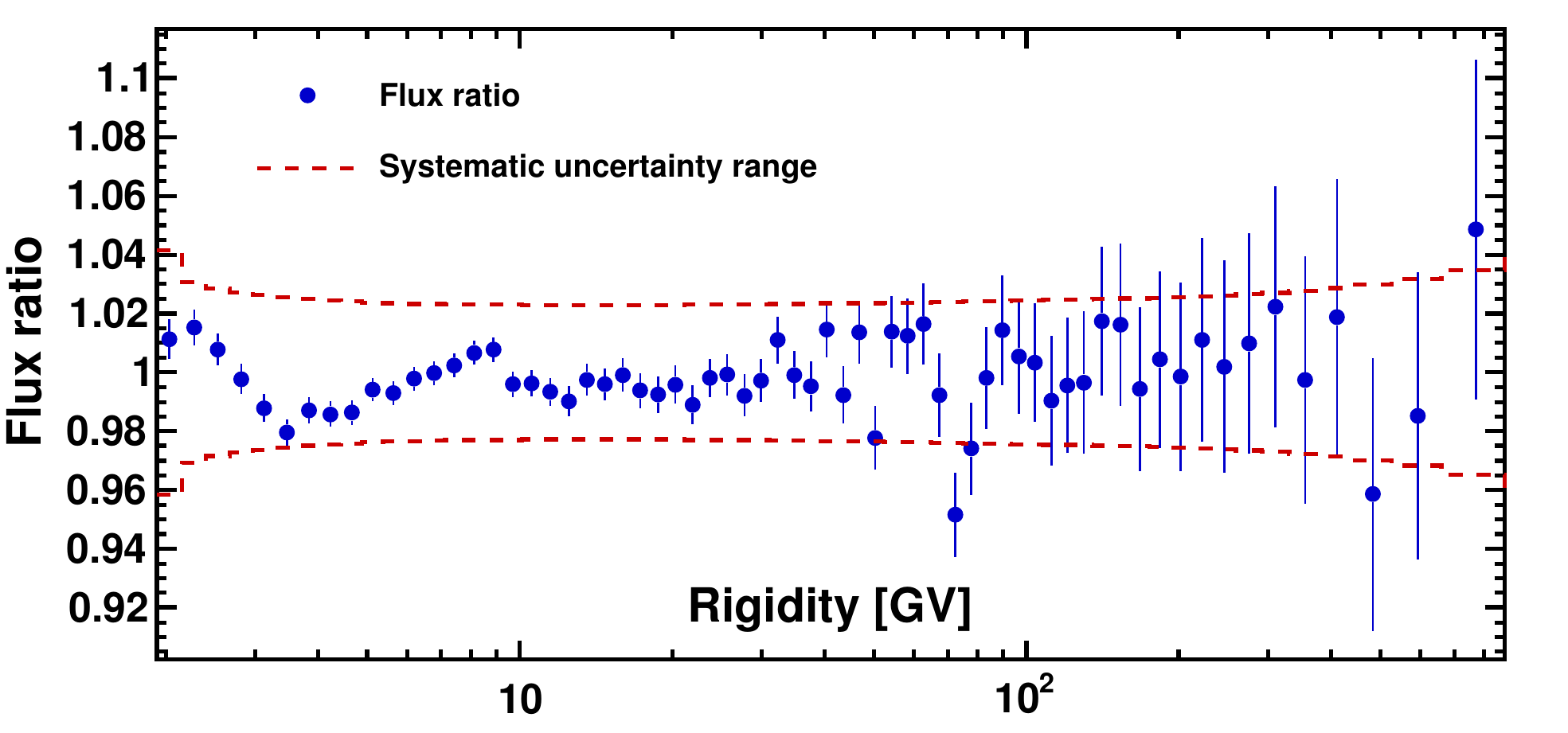}
\caption{Variation of the flux ratio measured using tracker L1 to L8 vs the full tracker (L1 to L9). The curves indicate the corresponding systematic errors.}
\label{fig:fig3}
\end{figure}

An additional verification was performed to ensure that the treatment of systematic errors is correct. Figure~\ref{fig:fig3} shows the ratio of two measurements of the carbon flux from 1.9 GV to 0.88 TV obtained using events passing through L1 to L8 and using events passing through L1 to L9. The good agreement between the two measurements verifies a) the systematic errors from unfolding, due to the difference in the resolution functions and, b) systematic errors from the acceptance, due to the difference in the geometric factor and the amount of material traversed.

Most importantly, several independent analyses were performed on the same data sample by different study groups. The current results of those analyses are consistent with the ones presented here.

\section{Results}
The current statistics of the carbon flux measurement above 200 GV are not enough to distinguish between a single power law or double power law~\cite{bib:ams-he} behavior at high energies. In the highest rigidity region, statistical error dominates.


In conclusion, precise knowledge of the carbon flux is important in understanding the origin, acceleration, and propagation of CRs. Our current measurement of the carbon flux from 1.9 GV to 2.6 TV is based on 8.3 million events with detailed studies of the systematic errors. At highest energies the current precision of the carbon flux measurement is limited by statistics.

%

\bigskip 
\bibliography{ECRS_proc_Yang_LI}
\bibliographystyle{rsc}


\end{document}